# Universal Intermediate Phases

# of Dilute Electronic and Molecular Glasses

J. C. Phillips

*Bell Laboratories, Lucent Technologies\*, Murray Hill, N. J. 07974-0636*

Generic intermediate phases with anomalous properties exist over narrow composition ranges adjacent to connectivity transitions. Analysis of both simple classical and complex quantum percolation shows how topological concepts can be used to understand many mysterious properties of high temperature superconductors, including the remarkably similar phase diagrams of $La_{2-x}Sr_xCuO_4$ and $C_{60}^{+y}$.

Ceramic high-temperature superconductors (HTSC) exhibit high $T_c$ 's over a narrow range of compositions adjacent to an insulator-metal phase transition (MIT). With a dopant density n one supposes that the density of *mobile* charge carriers, $n-n_{c1}$, tends continuously to zero at the connectivity MIT, while the average spacing between these carriers becomes very large. Whatever the screened interaction between mobile carriers may be, their interactions should become weaker and thus $T_c$ should tend gradually to zero, which is the opposite of the observed rapid collapse. In addition to the MIT there is generally a *second* transition at larger $n = n_{c2}$ from the intermediate superconductive phase, which has anomalous transport properties even in the normal state, to a normal metal (Fermi liquid) that is not superconductive, much like a lighter alkali metal. The second transition is unexpected and is first-order.

It seems that these *two* phase transitions cannot be explained by the effective medium approximation (EMA). The EMA is a high-density approximation that neglects fluctuations, and is intrinsically unsuitable in the low-density limit near phase transitions where $n-n_c$ is small and fluctuations are large[1,2]. In this limit for attractive interactions the fluctuations can become coherent and often belong to a *generic topological class* here identified and explained in the context of numerous examples.

Phase transitions in the simplest disordered off-lattice cases, partially quenched *classical* molecular systems, cannot be treated analytically. In several low-density molecular cases fluctuations have been simulated numerically, and even studied experimentally in considerable detail. The electronic *quantum* cases are still more difficult. They cannot be simulated numerically[3], in agreement with the comment[4] that "Some of the deepest problems in physics still surround the dynamics of glasses". The scaling behavior of nearly randomly doped semiconductor MIT's has been measured[5] in the limit $T \to 0$. These new developments share several generic features, most notably *an intermediate phase with anomalous properties remarkably similar to those observed in HTSC*.

In the dilute limit with attractive interactions, supercooled liquids and glasses may behave quite differently from crystals. In crystals the upper dimensionality where mean-field theory (Gaussian fluctuations) becomes valid is 2d in (**r**,**p**) classical models, and d +1 in (**p**,t) quantum models. In spin glasses the residual (low excess entropy) relaxation rate reaches an effective dimensionality $d^* = 1$ for both $d = 3$ and $d = 4$, and is probably independent of d, as sample size increases[6,7]. Thus at low densities attractive interactions can lead to dimensional collapse. Note, however, that because of *transverse* interactions, the bulk physics of one-dimensional subspaces embedded in d > 1 dimensions is *not the same* as that of isolated one-dimensional spaces. There are many theoretical studies of strictly d = 1 models, but the intermediate bulk phase discussed here is *not obtainable in these models*.

In ordinary electrolytes, with repulsive solute-solute ionic monopole interactions, liquid–gas phase transitions, as well as phase separation, occur, much as in hard-sphere fluids. With attractive dipolar interactions at *high* densities there is the usual liquid-gas transition, the condensed state having ferroelectric order. At *low* densities, however, there is no liquid-gas transition; the dipoles simply form curvilinear head-to-tail chains[8] that are at least metastable at low T. In dipolar solutions the EMA is valid for high densities, but it does not predict the low-density dimensionally collapsed daisy chains.

Network glasses also contain many surprises in the low-density limit. Here instead of the MIT one finds an elastic stiffness transition[9,10] that has been studied with considerable success both in numerical simulations[11] and in experiment[12]. This transition is analogous to the electronic MIT, and it occurs when the number of bond-stretching and bond-bending Lagrangian constraints per atom, $N_c$, equals d, the number of degrees of freedom per atom. Thus by low density of excess constraints one means the difference $\varepsilon = (N_c - d)/d \ll 1$; at such densities one finds, in numerical simulations, bulk percolating, curvilinear, rigid, topologically one-dimensional d* = 1 backbones[11].

The most striking new discovery in glass science in many decades has occurred in measurements of the kinetics of the glass transition using Modulated Differential Scanning Calorimetry[12]. These studies, which correlate extremely well with extensive elasticity and microscopic Raman scattering measurements, separate the enthalpy of the transition into reversing and nonreversing parts. As functions of composition or $\varepsilon$, the experiments have identified a $0 < \varepsilon < \varepsilon_2$ *reversibility window*, where the ratio of the nonreversing to the reversing transition enthalpy has dropped by a factor of 10 or more below typical values. This reduction reflects a lack of vortex entanglement (absence of rings) in unstressed backbones, and it identifies an intermediate phase associated with connected network flexibility. The width of the window, as well as its shape, is easily understood in terms of local network topologies specific to particular alloys. Examples shown in Fig. 1 include broad and narrow windows, the narrow one of width $\varepsilon_2 \sim 0.005$.

Although the electronic quantum cases cannot be simulated numerically, the topological similarities to the molecular cases just discussed are pervasive and instructive. Phase transitions in semiconductor *impurity bands* can be used as a *benchmark to test all proposed theories of the MIT in HTSC*. The nature of the impurity band transitions is easily identified and characterized in 3-d samples where the impurities are nearly randomly distributed. Evidence for the formation of one-dimensional coherent filamentary paths is provided by the scaling exponent $\alpha$ for the conductivity $\sigma$ in the limit $T \rightarrow 0$: $\sigma \propto (n-n_{c1})^{\alpha}$. With a classical random resistor model, one obtains $\alpha \sim 1$. Quantum mechanically, with a constant scattering rate, one obtains $\alpha = d/2$. In the EMA one would have $d = 3$ or $\alpha = 1.5$. The observed value[5] is $\alpha = 0.50(1)$, corresponding to a much sharper transition. This strongly implies $d = d^* = 1$, or one-dimensional filaments. There are alternative explanations[13] for $\alpha = 1/2$, but the EMA with $d = 3$ and plane-wave basis functions has clearly failed. Historical footnote: the possibility of a bulk intermediate phase of the electronic MIT based on reduced dimensionality was discussed[14] for free electrons by Wigner in 1935.

The one-dimensional filamentary embedded model readily explains[13] the wholly unexpected second transition, from the intermediate phase to the Fermi liquid, which is observed in specific heat data on Si:P. This transition is first order, and it occurs when the *transverse* confinement energy of the filaments becomes too large, so that lower energy is achieved by reverting to the locally isotropic $d = 3$ Fermi liquid; this occurs when the mobile dopant density $n-n_{c1}$ becomes too high. Similarly, in network glasses, when the density $\varepsilon$ of excess constraints becomes too high, the network is entangled[11] and the normal nonreversibility of the glass transition is restored. In impurity bands the second "crowding" transition occurs around $n_{c2} \sim 2n_{c1}$. The rings that stress the backbone of the network glass turn into branched filaments in the electronic case. Branching causes the local electronic one-dimensional filamentary wave packet phase variable to lose its meaning, and one reverts to the incoherent EMA. There the random phase approximation is often useful, because there are no separate phase variables as there are no isolated subspaces.

To understand the phase diagrams of HTSC one must allow for several additional factors that are specific to perovksite and pseudoperovksite oxides. All these materials are ferroelastic, and consequently are subject to large internal stresses that lead[15] to formation of orthorhombic (or other low symmetry) domains with dimensions of order 3 nm. In most cuprates the electrically active dopants are oxygen atoms or vacancies, and these have high mobilities. As a result, a new and very favorable mechanism develops that aids the formation of one-dimensional filaments: the dopants, which are disordered at high temperatures, can diffuse to form metallic chains, very similar to classical dipolar daisy chains. These metallic chains are embedded in an otherwise semiconductive environment, with semiconductive barriers (or pseudogaps) located in the nanodomain walls. Such chains will have very high conductivities even in the normal state, as the transverse ground state is non-degenerate. (For the same reason photons propagating in optical fibers have little attenuation.) The formation of such metallic chains maximizes the sample conductivity and leads to better screening of fluctuating ionic (internal) electric fields. This improved screening lowers the total energy, even at high annealing temperatures, where oxygen mobilities are high. Thus the dopants' preferred and kinetically accessible configuration (lowest free energy) is filamentary. The zigzag structure can be said to be topologically self-organizing, although there is no long-range order detectable by diffraction.

Such self-organization has many consequences. The electronic resistivity and Hall numbers in the normal state become linear in T above the pseudogap temperature, as suggested long ago[16], because the zigzag filamentary states form a narrow dopant band of high mobility states with non-perturbatively reordered energies pinned to $E_F$. To avoid planar nanodomain walls, the filaments must funnel carriers through resonant tunneling centers in the semiconductive planes that alternate with cuprate or other metallic planes. At such resonant tunneling centers the electron-phonon interactions become very large, due to reduced screening. (In the Si:P impurity band case an enhancement factor of 25 is obtained experimentally from the specific heat[13].) This resolves the paradox stated at the beginning of this paper: by concentrating the few carriers in even fewer filaments, not only do we avoid having diluted weak interactions, but because of poor screening at resonant tunneling centers, the funneled interactions are concentrated and thus greatly enhanced, accounting for the high values observed for $T_c$ in the cuprates.

The percolative model enables us to understand nanoscale phase separation, as shown[17] for LSCO ($La_{2-x}Sr_xCuO_4$) in Fig. 2. The *two* phase transitions seen explicitly in the electronic filling factor $f(x)$ are coupled to internal strains in the host lattice. The overall composition scale is expanded for $T_c(x)$ because of partially quenched phase overlap. The *two* phase transitions of the filling factor $f(x)$ become *two* phase immiscibility domes in $T_c(x)$, identified by the *two* pairs of *sharp breaks in slope* of $T_c(x)$ as it crosses spinodal boundaries.

*Sharp breaks in slope* are also visible in the phase diagrams[18] of $C_{60}^{+y}$, and these carry over to lattice-expanded $C_{60}^{+y}/ CH(Cl,Br)_3$. Although $T_c^{max}$ changes by more than a factor of 2 due to lattice expansion, the electronic spinodal regions of all three $C_{60}$-based phase diagrams closely resemble the phase diagram of under- and optimally-doped LSCO, with $y_{11} = 1.7$ and $y_{12} = 2.2$ holes/$C_{60}$. (Note that the linear spinodal regions all extrapolate to $T_c = 0$ at $y_{10} = 1.55(2)$, which should define the first MIT [without "tails"].) *Chains* of $C_{60}^{+z}$ spits will have lower energies and greatly enlarged e-p interactions due to enhanced screening of charge fluctuations that occur in space-filling $C_{60}^{+w}$ host molecules *transverse* to the $C_{60}^{+z}$ chains[19]. Spheroidal Bucky ball $C_{60}$ is deformed into a prolate ellipsoidal $C_{60}^{+z}$, leading to larger intermolecular overlap in the aligned $C_{60}^{+z}$ chains. These factors increase $T_c$ in $C_{60}^{+z}$ compared to $C_{60}^{-y}$, and suggest that e-p interactions, enhanced by dopant chain formation, are the cause of HTSC in *both* fullerites and cuprates. Moreover, the unexpected universality of $y_{max} = 3$ ($T_c^{max} = T_c(y_{max})$) between 3-fold orbitally degenerate electrons and 5-fold orbitally degenerate holes, for both chemically-doped and field-doped samples[18], is an automatic consequence of filamentary topology, as is the almost exact similarity between fullerene and $CaCuO_2$ phase diagrams[19,20].

This discussion has emphasized the phase diagrams of HTSC because these exhibit the largest degree of self-organization, and the most detailed anomalies. Disordered metal films with much lower $T_c$'s, such as (Si,Au), exhibit similar, but less detailed, correlations[21] between superconductivity and the MIT. The connections between the MIT, generic chain-based intermediate phases, and the many anomalies observed in HTSC, *both cuprate and fullerene,* have close topological parallels in more easily understood chain systems with much simpler internal chemistry. A discrete topological approach, rather than the continuum EMA, is likely to provide the most successful platform for understanding these mysterious materials.

*Retired. jcphillips8@home.com

# REFERENCES

1. These ideas are not new, but in most theories the failure of the EMA has been ignored because it appears to be impossible to analyze the nature of the fluctuations using brute-force methods. See R. Landauer, in *Electrical Transport and Optical Properties of Inhomogeneous Media*, American Institute of Physics Conference Proceedings, eds. J. C. Garland and D. B. Tanner (Am. Inst. Phys., New York, 1978), Vol. 40, pp.2-43.

2. J. C. Phillips, Physica C **252**, 188 (1995).

3. S. Lloyd, Phys. Rev. Lett.**71**, 943 (1993). Even the classical network case is insoluble for many particles, as it is related to the traveling salesman problem.

4. P. W. Anderson, Proc. Natl. Acad. Sci. USA **92**, 6653 (1995).

5. K.M. Itoh, M. Watanabe, Y. Ootuka, and E. E. Haller, Ann. Phys. (Leipzig*)* **1**, 1 (1999). The difference between this 3-d case and the "2-d" MOSFET case is that there the dopants are separated from the carriers, which results in loss of coherence in puddles between saddle points: Y. Meir, Phys. Rev. B **6307**, 3108 (2001).

6. I. A. Campbell and L. Bernardi, Phys. Rev. B **50**, 12643 (1994).

7. J. C. Phillips, Rep. Prog. Phys.**59**,1133 (1996).

8. D. Levesque and J. J. Weis, Phys. Rev. E **49**, 5131 (1994); T. Tlusty and S. A. Safran, Science **290**, 1328 (2000).

9. J. C. Phillips, J. Non-Cryst. Sol. **34**,153 (1979).

10. H. He and M. F. Thorpe, Phys. Rev. Lett. **54**, 2107 (1985).

11. M. F. Thorpe, M. V. Chubinsky, D. J. Jacobs and J. C. Phillips, J. Non-Cryst. Solids, **266-269**, 859 (2000).

12. D. Selvanathan, W. J. Bresser, P. Boolchand, and B. Goodman, Solid State Comm. **111**, 619 (1999); Y. Wang, J. Wells, D. G. Georgiev, P. Boolchand, K. Jackson, and M. Micoulaut, Phys. Rev. Lett. **87**, 185503 (2001).

13. J. C. Phillips, Sol. State Comm., **109**, 301 (1999).

14. M. Ross and L. H. Yang, Phys. Rev. B **64**, 134210 (2001). Most molecular quantum studies are carried out for metallic chain segments in vacuum, and so find the longer chains unstable against J-T dissociation into covalent dimers. These examples omit the gain in transverse screening energy for chains embedded in a strongly polarizable host, which produces an *anti-J-T effect*: J. C. Phillips, Phys. Rev. B **47**, 14132 (1993).

15. J. C. Phillips and J. Jung, Phil. Mag. B **81**, 745 (2001).

16. H. L. Stormer, A. F. J. Levi, K. W. Baldwin, M. Anzlowar, and G. S. Boebinger, Phys. Rev. B Solid State **38**, 2472 (1988).

17. J. C. Phillips, Phil. Mag. B (in press); LANL cond-mat/0106582.

18. J. H. Schon, Ch. Kloc, and B. Battlogg, Nature **408**, 549 (2001); Science **293**, 2432 (2001). These authors note that valence band narrowing enhances $T_c$ in $C_{60}^{+y}$.

19. J. H. Schon, M. Dorget, F. C. Beuran, X. Z. Zu, E. Arushanov, C. Deville Cavellin, and M. Lugues, Nature **414**, 434 (2001).

20. J. C. Phillips, LANL cond-mat/0112028.

21. M. S. Osofsky, R. J. Soulen, J. H. Claasen, G. Trotter, H. Kim, and J. S. Horwitz, Phys. Rev. Lett. **87**, 197004 (2001).

# FIGURE CAPTIONS

Fig. 1. The reversibility window in the enthalpy of the glass transition near the $\varepsilon = (N_c - d)/d = 0$ stiffness transition in network glasses. When the network stiffness is dominated by dead ends (disordered one-fold coordinated I atoms [12]), the window is narrow and sharp. In $Si_xSe_{1-x}$ linear chains are dominant, and then the window edges are still sharp, but the window is at least 20 times wider[12]. The two marked critical points for the wide window were determined independently from Raman scattering.

Fig. 2. Percolative interpretation[17] of data on the filling factor f(x), measured by Meissner volume and specific heat jump $\Delta C_p(x)$, and $T_c(x)$ in $La_{2-x}Sr_xCuO_4$. *Two* phase transitions in f(x) are reflected in *two* spinodal immiscibility domes in $T_c(x)$, indicated by the dotted lines. The first spinodal dome has also been observed[18] in $C_{60}^{+y}$, both pure *and* lattice-expanded.

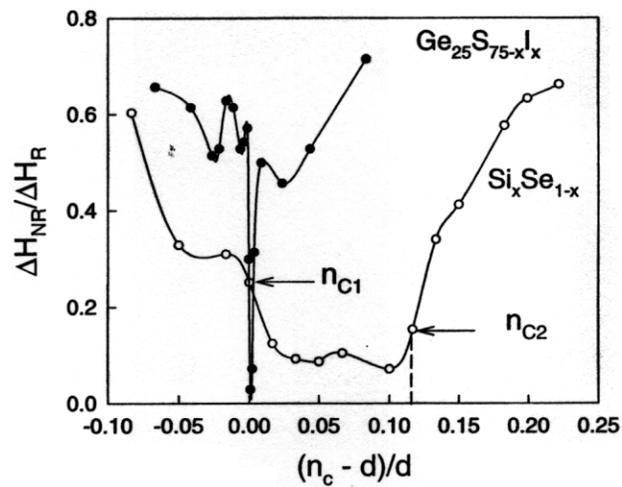

Fig. 1

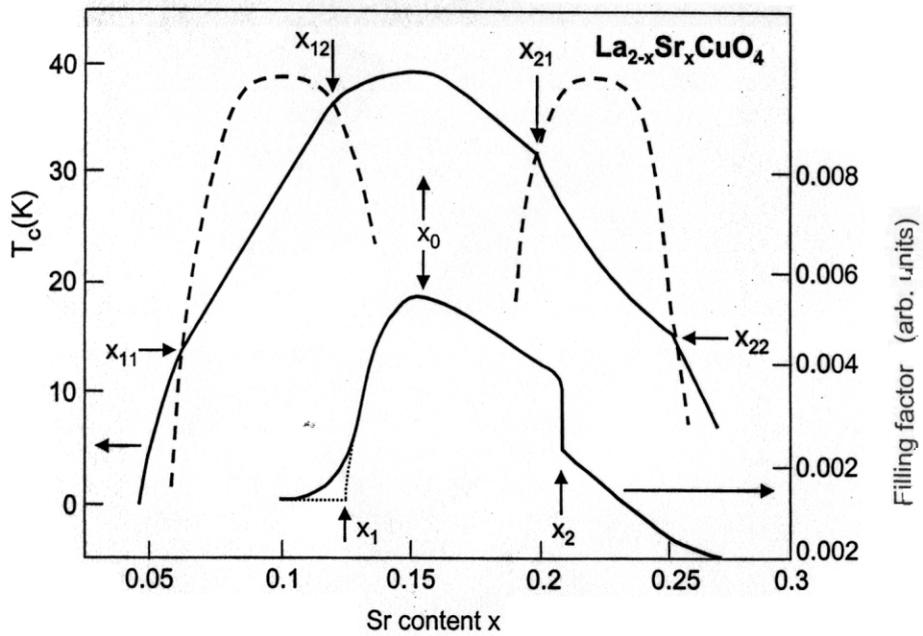

Fig. 2